\documentclass[11pt,a4paper]{article}
\usepackage{jcappub}

\usepackage{epsfig}
\usepackage{amsmath}
\usepackage{amssymb}

\title{
Enhanced neutrino signals from dark matter annihilation in the Sun via metastable mediators 
}
\author{Nicole F. Bell and Kalliopi Petraki}
\affiliation{School of Physics, The University of Melbourne, Victoria 3010, Australia}

\emailAdd{n.bell@unimelb.edu.au}
\emailAdd{kpetraki@unimelb.edu.au}

\date{\today}

\abstract{
We calculate the neutrino signal resulting from annihilation of
secluded dark matter in the Sun. In this class of models, dark matter
annihilates first into metastable mediators, which subsequently decay
into Standard Model particles.  If the mediators are long lived, they
will propagate out from the dense solar core before decaying.
High energy neutrinos undergo absorption in the Sun.  In the standard
scenario in which neutrinos are produced directly in the centre of the
Sun, absorption is relevant for $E \gtrsim 100$ GeV, resulting in 
a significant suppression of the neutrino spectrum
beyond $E \sim 1$ TeV.
In the secluded dark matter scenario, the neutrino signal is greatly
enhanced because neutrinos are injected away from the core, at lower
density.  Since the solar density falls exponentially with radius,
metastable mediators have a significant effect on the neutrino flux,
even for decay lengths which are small compared to the solar radius.
Moreover, since neutrino detection cross sections grow with energy,
this enhancement of the high energy region of the neutrino spectrum
would have a large effect on overall event rates.
}
\begin{document}
\maketitle
\newcommand{\nc}{\newcommand}
%----------------------------------------------
%%% Languages
\nc{\sen}[0]{\selectlanguage{english}}
\nc{\sgr}[0]{\selectlanguage{greek}}
%----------------------------------------------
%%% Typeface
\def\cal{\mathcal}
\def\rm{\mathrm}

\nc{\eq}[1]{Eq.~(\ref{#1})}
\nc{\eqs}[1]{Eqs.~(\ref{#1})}

\nc{\pd}{\partial}
%----------------------------------------------
%%% Spacing
\def\bs{\bigskip}
\def\ms{\medskip}
\def\sms{\smallskip}
%----------------------------------------------
%%% Environments
\nc{\beq}{\begin{equation}}
\nc{\eeq}{\end{equation}}
\nc{\bea}{\begin{eqnarray}}
\nc{\eea}{\end{eqnarray}}
\def\bal{\begin{align}}
\def\eal{\end{align}}
\nc{\bit}{\begin{itemize}}
\nc{\eit}{\end{itemize}}
\nc{\benu}{\begin{enumerate}}
\nc{\eenu}{\end{enumerate}}
\nc{\bdes}{\begin{description}}
\nc{\edes}{\end{description}}
%----------------------------------------------
%%% Math-mode typesetting shortcuts

\nc{\nn}{\nonumber}

\nc{\sub}[1]{_{\rm{#1}}}
\nc{\ssub}[1]{_{_\rm{#1}}}
\nc{\super}[1]{^{\rm{#1}}}
\nc{\ssuper}[1]{^{^\rm{#1}}}

\nc{\slashed}[1]{{#1}\hspace{-2mm}/}

\nc{\pare}[1]{\left( #1 \right)}
\nc{\sqpare}[1]{\left[ #1 \right]}
\nc{\ang}[1]{\langle #1 \rangle}
\nc{\abs}[1]{\left| #1 \right|}

%%%% Dirac matrix symbols
\def\g5{\gamma_{5}}

%%%% Units
\def \eV{\: \rm{eV}}
\def\keV{\: \rm{keV}}
\def\MeV{\: \rm{MeV}}
\def\GeV{\: \rm{GeV}}
\def\TeV{\: \rm{TeV}}

\def\erg{\: \rm{erg}}

\def \cm{\: \rm{cm}}
\def \km{\: \rm{km}}
\def \pc{\: \rm{pc}}
\def\kpc{\: \rm{kpc}}
\def\Mpc{\: \rm{Mpc}}
\def\Gpc{\: \rm{Gpc}}
\def\AU{\: \rm{A.U.}}

\def\sr{\: \rm{sr}}

\def \snd{\: \rm{s}}
\def  \yr{\: \rm{yr}}
\def \Myr{\: \rm{Myr}}
\def \Gyr{\: \rm{Gyr}}

%%%% Greek Letters
%%%%% lowercase
\def\a{\alpha}
\def\b{\beta}
\def\g{\gamma}
\def\d{\delta}
\def\e{\epsilon}
\def\z{\zeta}
\def\h{\eta}
\def\th{\theta}
\def\i{\iota}
\def\k{\kappa}
\def\l{\lambda}
\def\m{\mu}
\def\n{\nu}
\def\ks{\xi}
\def\om{o}
\def\p{\pi}
\def\r{\rho}
\def\s{\sigma}
\def\t{\tau}
\def\y{\upsilon}
\def\f{\phi}
\def\x{\chi}
\def\ps{\psi}
\def\w{\omega}

\def\ve{\varepsilon}
\def\vr{\varrho}
\def\vs{\varsigma}
\def\vf{\varphi}

%%%%% UPPERcase
\def\G{\Gamma}
\def\D{\Delta}
\def\Th{\Theta}
\def\L{\Lambda}
\def\Ks{\Xi}
\def\P{\Pi}
\def\S{\Sigma}
\def\Y{\Upsilon}
\def\F{\Phi}
\def\Ps{\Psi}
\def\W{\Omega}

%----------------------------------------------      % personal LaTeX macros

\section{Introduction \label{sec intro}}

The gravitational capture and subsequent annihilation of dark matter
(DM) in the Sun offers a compelling possibility for DM
detection.  DM particles passing through the Sun are expected
to scatter off nuclei, lose energy, and become trapped by the
gravitational field. Multiple scatterings cause DM to sink towards the
centre of the Sun. If DM self-annihilates, the capture is
quickly balanced by the annihilation of DM particles. The intensity of
the annihilation signal is then a probe of the DM scattering cross
section on nucleons~\cite{Press:1985ug,Gould:1987ir}.
Signals from DM capture in the Sun (or the Earth) are expected to be
detectable in the future. Standard Model (SM) particles other than
neutrinos, produced in the annihilations, interact strongly with the
interior of the Sun and are largely absorbed.  In this process though,
they produce high-energy neutrinos which escape, and can be
potentially seen by neutrino detectors, such as IceCube and DeepCore.
Extensive studies of the expected neutrino signal from DM annihilation
via various SM channels have been performed, in a model-independent
way~\cite{Cirelli:2005gh,Blennow:2007tw}, and for particular
models~\cite{Kamionkowski:1991nj,Crotty:2002mv,Baer:2004qq,Barger:2001ur,Berezinsky:1996ga,Bergstrom:1996kp,Bergstrom:1998xh,Bergstrom:1997tp,Bertin:2002ky,Bertin:2002sq,Bottino:1991dy,Bottino:1998vw,Faraggi:1999iu,Feng:2000zu,Gelmini:1990je,Halzen:1991kh}.
Limits from Super-Kamiokande and IceCube on annihilation of WIMP and Kaluza-Klein DM in the Sun and the Earth have been reported in
Refs.~\cite{Desai:2004pq,Abbasi:2009uz,Abbasi:2009vg}.

It is possible, however, that DM does not annihilate directly into SM
particles, but rather into metastable mediators which subsequently
decay into SM states, $\chi\chi \rightarrow VV \rightarrow
\textrm{SM}$, as recently discussed in~\cite{Pospelov:2007mp,Finkbeiner:2007kk, ArkaniHamed:2008qn,Pospelov:2008jd,Rothstein:2009pm,Chen:2009ab}.  
In such models, the thermal relic WIMP DM scenario can be realised as usual, while
there is also the potential to explain astrophysical observations,
e.g. the positron excess observed by
PAMELA~\cite{ArkaniHamed:2008qn,Pospelov:2008jd}.  The seclusion of DM
from the SM, which in this class of models communicate only via
metastable mediators, can dramatically change the annihilation
signature of DM captured in the Sun. For example, if the
mediators are sufficiently long-lived to escape the Sun before
decaying, they can produce detectable charged-particle or $\g$-ray
fluxes, as discussed in
Refs.~\cite{Batell:2009zp,Schuster:2009fc}.  

If the mediators are short-lived and decay in the interior of the Sun,
energetic neutrinos remain the only signature.  However, even for
decay of $V$ inside the Sun, the neutrino signal can be dramatically
enhanced compared to the standard scenario.  This is because high
energy neutrinos can interact with nuclei and be absorbed before
escaping the Sun.  In the standard scenario, in which neutrinos are
produced at the centre of the Sun, the neutrino energy spectrum is
damped as $e^{-E_\nu/\cal{E}}$, with a critical energy of $\cal{E}
\sim 100$ GeV.  However, the solar density decreases exponentially
with radius, so if neutrinos are injected by $V$ decay at larger radii
they are subject to much less absorption as they propagate out of the
Sun.  We thus expect the critical energy scale for absorption to
increase exponentially with the neutrino injection radius, increasing
rapidly once the injection point is moved outside the dense core.

For simplicity, we will consider the case where the mediators, $V$, decay
directly to neutrinos $V\rightarrow \nu \bar{\nu}$, as this will
suffice to illustrate the difference between DM annihilations with and
without mediators.  It is of course plausible that the mediators decay
to other SM model particles, which subsequently produce energetic
neutrinos.  Although we shall not study this case in detail, we note
that greatly enhanced signals are also expected in this scenario.  For
instance, if DM annihilates directly to light quarks or muons at the
centre of the Sun, the neutrino yield is negligible, as these
particles quickly lose energy and are absorbed before they can decay
to produce neutrinos.  However, if a mediator decay were to inject
light quarks or muons at radii beyond the dense core, then a
non-negligible neutrino flux would result; absorption effects are of
course completely avoided if the decay occurs outside the solar
radius~\cite{Batell:2009zp}.

In this paper, we consider the neutrino signal at the Earth from
annihilation of secluded DM in the Sun, for a range of mediator
lifetimes and DM masses.  In Sec.~\ref{sec prop} we describe the
production of neutrinos from decay of the mediator, review the
relevant aspects of neutrino interactions in the Sun, namely
absorption, regeneration, and flavour evolution.  We also describe the
numerical approach we follow to calculate the neutrino signal at the
Earth.  Our main results are presented in Figs~\ref{All tau e} --
\ref{10TeV}.  In Sec.~\ref{sec signal} we discuss the results, and
compare them with the standard case of DM annihilation directly into
SM particles.

%%%%%%%%%%%%%%%%%%%%%%%%%%%%%%%%%%%%%%%%%%%%%%%%%%%%%%%%%%%%%%%%%%%%
%%%%%%%%%%%%%%%%%%%%%%%%%%%%%%%%%%%%%%%%%%%%%%%%%%%%%%%%%%%%%%%%%%%%

\section{Neutrino production and propagation in the Sun \label{sec prop}}

During the capture and thermalisation of dark matter in the Sun, the
DM particles undergo multiple scatterings and become concentrated
within a rather small region of size $\sim 0.01 R_\odot
\sqrt{100 \GeV/m_\x}$ around the centre of the Sun, where $R_\odot$ is
the solar radius.  The finite size of this region has a negligible
effect on the final neutrino
spectra~\cite{Press:1985ug,Cirelli:2005gh,Blennow:2007tw}, and thus
the DM annihilations can be assumed to take place in the centre of the
Sun.

In the scenario considered here, DM annihilates into metastable mediators
\beq
\x\x \rightarrow V V, \label{annih}
\eeq
which subsequently decay into neutrino pairs
\beq
V \rightarrow \n \bar{\n}.
\label{decay}
\eeq
The mediators are emitted with Lorentz factor $\g = m_\x/m_V$.  In
the rest frame of the Sun, the neutrinos produced from mediator decay
uniformly span the energy interval
\beq
\frac{1}{2} (1-\b) \:  m_\x \leqslant E_\n \leqslant \frac{1}{2} (1+ \b) \: m_\x,
\label{E nu}
\eeq
where $\b$ is the mediator velocity, and have an angular dispersion $\d\th \sim 1/\g$ around the momentum axis of the mediator~\cite{Batell:2009zp}. As long as $\g \gg 1$, the neutrinos are effectively emitted radially outwards, with energies $E_\n \leqslant m_\x$. This allows us to use 1-dimensional evolution equations, similarly to the approach commonly adopted in the standard scenario of DM decay directly into SM particles~\cite{Cirelli:2005gh,Blennow:2007tw,Crotty:2002mv}.

We will focus on DM masses $100 \GeV \lesssim m_\x \lesssim 10
\TeV$. Neutrinos of energies lower than 100~GeV have negligible
interactions in the interior of the Sun, and their final spectrum is
not very different from their spectrum at production.  At $E_\n
\gtrsim 1 \TeV$, the absorption of neutrinos in the Sun becomes quite
severe, strongly suppressing the signal if neutrinos are injected in
the centre of the Sun, as is the case for the standard scenario of DM
annihilation directly into SM particles.  However, if DM annihilates
into metastable mediators, which travel some distance before decaying,
the neutrinos transverse smaller optical depth in the Sun and are thus
absorbed less. As we shall see, this can change the total flux and
spectral shape of the signal for any DM mass $m_\x \gtrsim 100 \GeV$,
and more dramatically so for $m_\x \gtrsim 1 \TeV$.

Once neutrinos are produced, they undergo charged and neutral current
scattering with nuclear matter in the Sun, which results in both
absorption of high energy neutrinos, and re-injection of neutrinos at
lower energy.  They also undergo flavour oscillations.  For the energy
range of interest, the neutrino flavour oscillations and their
scattering interactions decouple. The matter potential in the Sun
ensures negligible mixing between $\n_e$ and $\n_\m, \n_\t$ until the
point of the MSW resonance. For $5 \GeV \lesssim E_\n \lesssim 10
\TeV$, which encompasses our energy range of interest, the resonance
occurs after interactions have become unimportant, as will be shown in
Sec.~\ref{sec osc}.  
(Neutrinos below 5 GeV make an insignificant contribution to the
signals we consider.)
On the other hand, $\n_\m$ and $\n_\t$ mix
maximally, with oscillation length typically smaller or comparable to
the interaction length. It is then a good approximation to assume that
oscillations rapidly equidistribute the $\n_\m$ and $\n_\t$ fluxes, and thus
identify their densities and average over their interactions.

The above considerations allow us to follow the evolution of the
flavour eigenstates individually, rather than using the full
multi-flavour density-matrix formalism. The evolution equations have
the form
\beq
\frac{\pd \rho_j}{\pd r}  =  \left.\frac{\pd \r_j}{\pd r}\right|\ssub{inj} + \left.\frac{\pd \r_j}{\pd r}\right|\ssub{NC} + \left.\frac{\pd \r_j}{\pd r}\right|\ssub{CC},
\label{prop}
\eeq
where $\r(r,E) dE$ is the neutrino flux, and $j=e$ or $\m,\t$, for the electron and the averaged muon and tau neutrino fluxes respectively. Similar equations hold for the antineutrinos.
The first term on the right-hand side corresponds to the neutrino
injection from the mediator decay. The second and the third terms
describe the neutrino scattering due to neutral-current (NC) and
charged-current (CC) interactions. Neutrino-neutrino scatterings are
unimportant, since the neutrino densities are small, and thus the
evolution equations remain linear. Each term in Eq.(\ref{prop}) is
described in detail in the following subsections.

In calculating the neutrino propagation inside the Sun, we adopt the solar nucleon and electron density profiles from the Standard Solar Model~\cite{Bahcall:2004pz}.
For $r \lesssim 0.9 R_\odot$, the nucleon density can be well approximated by
\beq
N_S(r) \simeq N_0 e^{-\frac{r}{\k R_\odot}},
\label{sun}
\eeq
with  $N_0 = 1.3 \times 10^{26} \cm^{-3}$ and $\k=0.1$.
It will be useful to define the dimensionless optical depth (independently of the approximation of \eq{sun}):
\beq
x(r) \equiv \frac{1}{N_0 \: \k R_\odot} \int_0^r  N_S(r') dr',
\label{opt length}
\eeq
which can be inverted to give $r(x)$. For the region of validity of \eq{sun}, we obtain the analytic expressions
\beq
x_a(r) = 1-e^{-\frac{r}{\k R_\odot}}, \quad r_a(x)=\k R_\odot \ln\pare{\frac{1}{1-x}}.
\label{x,r anal}
\eeq

\subsection{Injection from mediator decay \label{sec inj}}

Neutrinos are injected by the decay of the metastable mediators
produced in the DM annihilations in the center of the Sun, as
described in \eqs{annih}, \eqref{decay}. In the steady state regime,
the annihilation balances the DM capture in the Sun, $\G\sub{ann} =
C_\odot/2$, where the capture rate is
approximately~\cite{Gould:1987ir,Kamionkowski:1991nj}
\beq
C_\odot \sim 10^{21} \snd^{-1} \pare{\frac{100 \GeV}{m_\x}} \pare{\frac{\s\sub{sc}}{10^{-42} \cm^2}}.
\label{capture}
\eeq

We will assume that the mediators decay with equal branching ratio to
each of the three neutrino flavours $\rm{BR}_\n \equiv \rm{BR}_{e,\m,\t}$. In the
rest frame of the Sun, the neutrino energy spectrum per $V$ decay is
flat, $f(E)dE = (\rm{BR}_\n / \b m_\x) dE$, for $(1-\b)/2 \leqslant E/m_\x
\leqslant (1+\b)/2$.  The neutrino injection rate is then
\beq
\left.\frac{\pd \r(r,E)}{\pd r}\right|\ssub{inj} = \frac{C_\odot \: \rm{BR}_\n}{\b m_\x} \frac{1}{\b\g\t} \exp{\pare{-\frac{r}{\b\g\t}}}.
\label{inj}
\eeq 
We take the lifetime, $\t$, of the mediators to be a free parameter,
subject only to the big bang nucleosynthesis (BBN) constraint, $\t < 1 \snd$~\cite{Chen:2009ab}.
In order to disentangle our results from any astrophysical
uncertainties and model-dependence, 
we shall present our results for the neutrino fluxes normalized to the product
$C_\odot \: \rm{BR}_\n$.

\subsection{Interactions \label{sec inter}}

Neutrinos in the Sun interact with the nuclei via neutral and charged currents. The interaction cross sections are rather insensitive to the proton to neutron ratio inside the sun, which varies from 2 in the center, to 7 in the outer regions. We thus adopt the constant value $p/n = 3$ throughout.  Scattering on electrons is suppressed in comparison to scattering on nuclei by the ratio of the electron to nucleon mass, $m_e/m_N$, and we will thus ignore it in this analysis.

\subsubsection{Neutral-current interactions \label{sec NC}}

The NC scatterings $\n_l N \rightarrow \n_l N$ and $\bar{\n}_l N \rightarrow \bar{\n}_l N$ shift the neutrino distributions towards lower energies. The process can be described as removal of a neutrino from the flux and re-injection at a lower energy, as follows
\beq
\left.\frac{\pd \rho}{\pd r}\right|\ssub{NC} =  N_S(r) \left[-\s\ssub{NC}(E) \r(r,E) + \int_E^\infty dE' \left.\frac{d \s\ssub{NC}}{dE}\right|_{E'\rightarrow E} \r(r,E') \right],
\label{NC}
\eeq
where $N_S(r)$ is the nucleon density of the Sun. The differential and total NC cross sections at the energy range of interest have the form~\cite{Gandhi:1998ri,Cirelli:2005gh,Crotty:2002mv}
\begin{alignat}{2}
\left.\frac{d\s\ssub{NC}}{dE}\right|_{E' \rightarrow E} &=  \frac{2 G_F^2 m_N}{\pi} \sqpare{a+b(E/E')^2}, \label{NC diff sigma} \\
\s\ssub{NC}(E)                                          &=  \frac{2 G_F^2 m_N}{\pi} (a+b/3) E.  \label{NC sigma}
\end{alignat}
NC scattering is flavour-blind, with
\beq
\begin{alignedat}{4}
a_\n         &\simeq 0.06; &\ b_\n         &\simeq 0.02;  \\
a_{\bar{\n}} &=      b_\n; &\ b_{\bar{\n}} &= a_\n.
\end{alignedat}
\label{a,b}
\eeq

\subsubsection{Charged-current interactions \label{sec CC}}

CC interactions convert a neutrino to an almost collinear charged lepton, $\n_l N \rightarrow l^- N'$ and $\bar{\n}_l N \rightarrow l^+  N'$.
For electron-type neutrinos, this corresponds to removal of a neutrino from the flux.
Muon-type neutrinos produce $\m^\mp$ which decay and re-inject neutrinos in the flux, 
but not before they have thermalised in the plasma and lost most of their energy. The 
re-injected neutrinos with energies $\sim \rm{few} \times 10 \MeV$ do not contribute 
to the high-energy neutrino signature of DM annihilations, and for this purpose muon 
neutrinos that interact via CC are considered absorbed.

Charged-current interactions of $\n_\t$ and $\bar{\n}_\t$ produce $\t^\mp$
leptons which decay promptly, before losing their energy, and re-inject
energetic neutrinos in the flux.  Besides re-injecting a tau-flavour
neutrino at a lower energy, the leptonic decays $\t^- \rightarrow X \n_\t$
($\t^+ \rightarrow X \bar{\n}_\t$) also produce a $\bar{\n}_e$
($\n_e$) $17.8\%$ of the time, and a $\bar{\n}_\m$ ($\n_\m$) $17.4\%$
of the time. (The $\t^\mp$ decay hadronically with probability $64.8\%$.)
Since the contribution to the $\bar{\n}_e$ ($\n_e$) and $\bar{\n}_\m$
($\n_\m$) fluxes from $\n_\t$ ($\bar{\n}_\t$) regeneration is
suppressed by the branching fraction $\sim 0.18$, we will ignore this
effect\footnote{As with absorption, regeneration has a smaller effect in the mediator scenario compared to the standard scenario, so this is an even better approximation than usual.}.
This stands in agreement with similar neutrino
signal calculations in various environments which have shown that such
a contribution is not significant~\cite{Dutta:2002zc}.

The CC absorption and regeneration is described by
\beq
\left.\frac{\pd \r_l}{\pd r}\right|\sub{CC} =  N_S(r) \left[ - \r_l(r,E) \s\ssub{CC}(E) +  \int_E^\infty dE' \r_l(r,E') f_{l \rightarrow l}(E,E') \right]. 
\label{CC}
\eeq
The CC inelastic scattering cross section does not depend on the
neutrino flavour, as long as we ignore the effect of the $\tau$ mass.
This is a good approximation since the $\tau$ mass is significant
only for $E \lesssim 100 \GeV$, and at those energies the neutrino
interactions in the Sun are unimportant.  The CC cross section
is~\cite{Gandhi:1998ri,Cirelli:2005gh,Crotty:2002mv,CooperSarkar:2007cv} \beq
\s\ssub{CC}(E) = \frac{2 G_F^2 m_N}{\pi} c E, \label{CC diff sigma}
\eeq
with
\beq
c_\n \simeq 0.19, \ c_{\bar{\n}} \simeq 0.13.
\label{c}
\eeq
For the electron and muon flavours we set $f_{l \rightarrow l}
\rightarrow 0$, as discussed above. For tau neutrinos these functions
encode the appropriate convolution of $\t^\mp$ spectrum arising from
CC $\n_\t, \bar{\n}_\t$ interactions, with the regenerated $\n_\t,
\bar{\n}_\t$ spectrum from the $\t^\mp$ decay.  They depend only
mildly on the incident neutrino energy, and are merely functions of
the ratio of the outgoing to incident neutrino energy, $E/E'$.  We
thus adopt the form of $f_{\t \rightarrow \t}(E/E')$ and $f_{\bar{\t}
  \rightarrow \bar{\t}}(E/E')$ as presented e.g. in
Ref.~\cite{Cirelli:2005gh} for incident energy $E'=400 \GeV$.

\begin{figure}[t]
\centering
\includegraphics[width=0.6\linewidth]{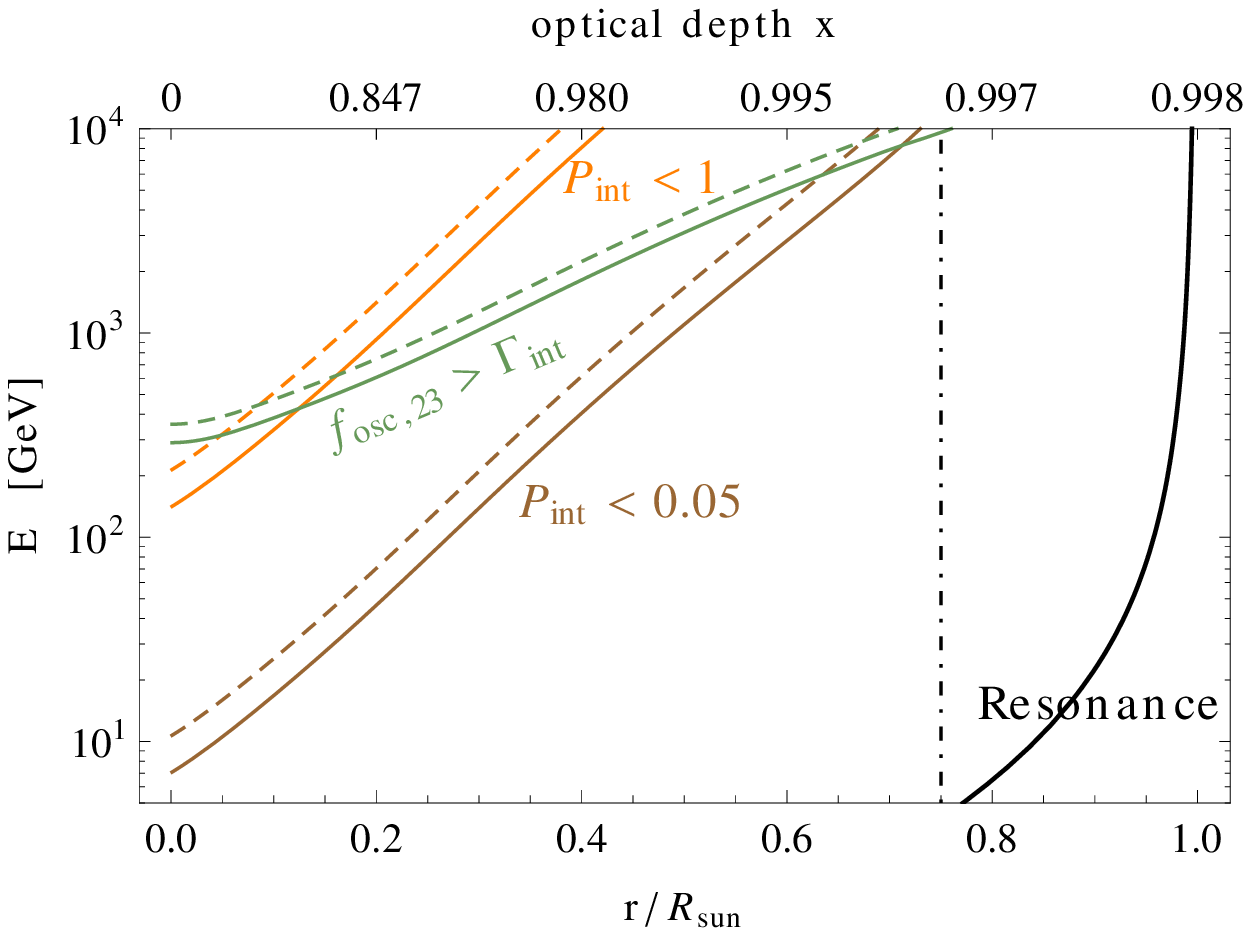}
\caption{Interaction probability contours, $P\sub{int} = 1$ (orange)
  and $P\sub{int}=0.05$ (brown), for neutrinos (solid lines) and
  antineutrinos (dashed lines).  The black line marks the point of the
  $\nu_e$--$\nu_{\mu,\tau}$ resonance. For the energy range of
  interest, interactions have effectively ceased when neutrinos
  undergo resonance.  We neglect interactions beyond the benchmark
  point $r_c=0.75 R_\odot$ (dot-dashed line).
Oscillations between muon and tau neutrinos occur faster than interactions for energies below the green lines. Neutrinos are then equidistributed between the two flavour eigenstates.}
\label{phase space}
\end{figure}

As will become clear in Sec.~\ref{sec prop sum}, the neutrino total interaction strength inside the Sun can be conveniently gauged in comparison to the energy scale
\beq
\cal{E}  \equiv \sqpare{\frac{2 G_F^2 m_N}{\pi} (a + b/3 +c) N_0 \: \k R_\odot}^{-1}.  \label{cal E}
\eeq
Using \eqs{a,b} and \eqref{c}, for neutrinos and antineutrinos respectively, we have
\beq
\cal{E}_\n  \simeq 140\GeV; \qquad \cal{E}_{\bar{\n}} \simeq 213 \GeV. \label{cal E num}
\eeq
The values of $\cal{E}$ in Eq.(\ref{cal E num}) indicate the
approximate energy scale at which interactions become a significant
effect and  illustrate that absorption is relevant at lower energies
for neutrinos than antineutrinos.
In terms of $\cal{E}$, the neutrino total interaction cross section becomes
\beq
\s\sub{tot}(E) = \frac{1}{N_0 \kappa R_\odot} \frac{E}{\cal{E}}.
\label{sigma tot}
\eeq
The neutrino mean free path, $\l_f$, defined by
\beq
\int_r^{r+\l_f} \s\sub{tot}(E) N_S(r') dr' =1, \nn
\label{mean free path}
\eeq
corresponds to optical depth 
\beq
\d x_f = \cal{E}/E,
\label{dx1}
\eeq
and the neutrino interaction probability is
\beq
P\sub{int} = \int_r^{R_\odot} \s\sub{tot}(E) N_S(r') dr' = \frac{E}{\cal{E}} (x_\odot-x),
\label{Pint}
\eeq
where $x_\odot \simeq 0.9975$ is the optical depth at $r=R_\odot$. In Fig.~\ref{phase space}, we sketch the interaction probability contours $P\sub{int} = 1$ and $P\sub{int} = 0.05$, on the $E \ vs \ r$ plane.

%%%%%%%%%%%%%%%%%%%%%%%%%%%%%%%%%%
%%%%%%%%%%%%%%%%%%%%%%%%%%%%%%%%%%%

\subsection{Flavour oscillations \label{sec osc}}

\subsubsection{$\n_e - \n_\m$ oscillations}

\begin{figure}[t]
\centering
\includegraphics[width=0.55\linewidth]{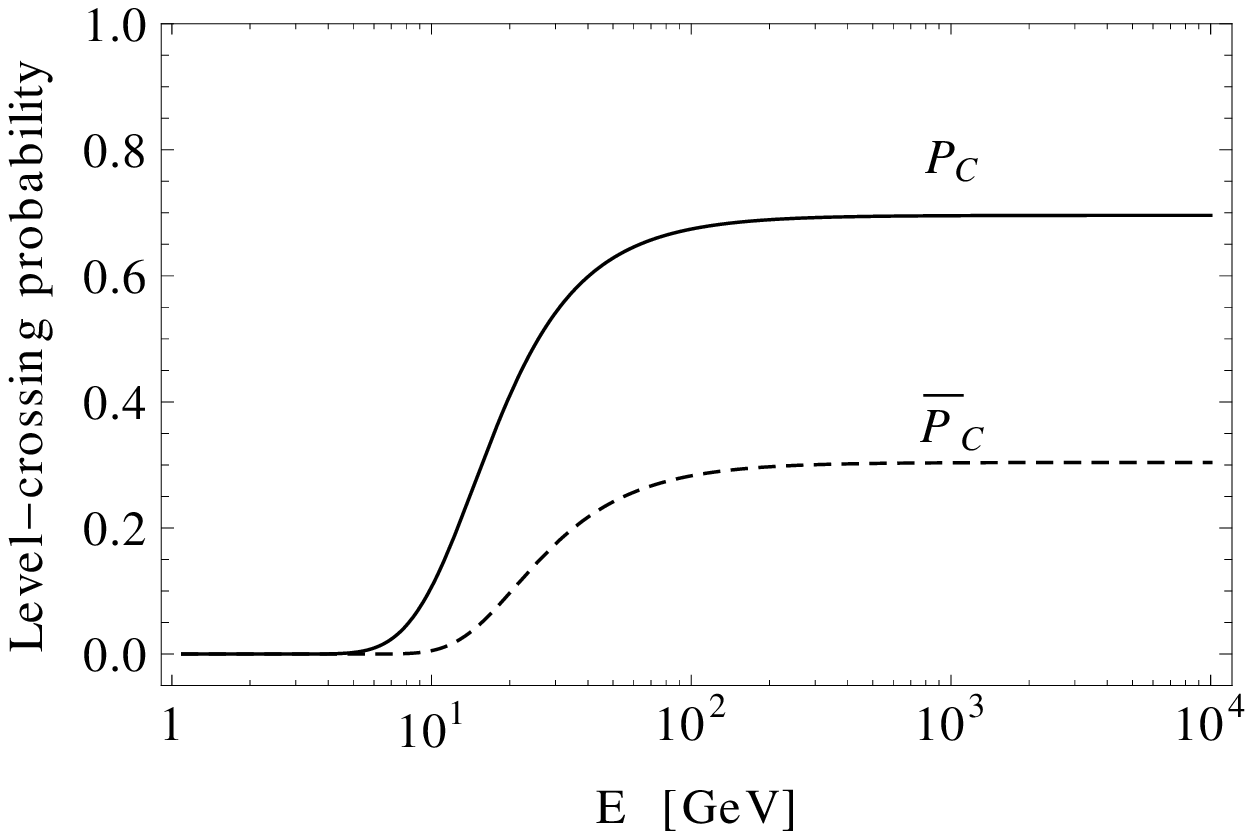}
\caption{Level-crossing probability between mass eigenstates 1 \& 2, for neutrinos (solid) and antineutrinos (dashed).}
\label{PC plot}
\end{figure}

In the inner region of the Sun, the neutrinos experience a matter
potential (or refractive index) which aligns the flavour eigenstates
with (effective) mass eigenstates, and suppresses oscillations between
electron and mu/tau neutrinos.  At lower densities, as the matter
potential decreases, the e--mu/tau mixing angle increases and the
neutrinos go through an MSW resonance.  The point of resonance occurs when
\beq
\frac{\D m^2_{21}}{2 E} \simeq \sqrt{2} \: G_F N_e(r),
\label{resonance}
\eeq
where $N_e(r)$ is the electron density profile of the Sun.  The radius
at which the resonance takes place is shown in Fig.~\ref{phase space}.
While low energy solar neutrinos go through this resonance
adiabatically, remaining in a given mass eigenstate, non-adiabatic
effects (in which neutrinos may flip to the other mass eigenstate) are
important for neutrinos with energies greater than several $\sim 10
\GeV$.  The level-crossing probability at the point of resonance 
$r=r\sub{res}(E)$ is (see e.g.~\cite{Strumia:2006db})
\beq
P_C(E) \simeq  \frac{e^{\tilde{\g} \cos^2\th_{12}} - 1}{e^{\tilde{\g}} - 1}, \qquad \tilde{\g}(E) = \frac{\p r_0 \: \D m^2_{21}}{E},
\label{PC}
\eeq
where
\beq
r_0 = \abs{\frac{d \ln N_e}{dr}}^{-1}_{r=r\sub{res}(E)},
\eeq
while for antineutrinos
\beq
\bar{P}_C = P_C \: [\th_{12} \rightarrow \pi/2 - \th_{12}].
\label{PCbar}
\eeq
The crossing probabilities $P_C(E)$ and $\bar{P}_C(E)$ are shown in Fig.~\ref{PC plot}; they become approximately constant for $E \gtrsim 100$ GeV.

%%%%%%%%%%%%%%%%%%%%%%%%%%%%%
%%%%%%%%%%%%%%%%%%%%%%%%%%%%

After resonance, the fluxes of the three mass eigenstates are
\beq
\begin{alignedat}{2}
\r_1   &= \r_e^{(i)} P_C      + \r_{\m,\t}^{(i)} (1-P_C), \\
\r_2   &= \r_e^{(i)} (1-P_C)  + \r_{\m,\t}^{(i)} P_C,  \\
\r_3   &=                       \r_{\m,\t}^{(i)},
\end{alignedat}
\label{flux1,2,3}
\eeq
where $\r_e^{(i)}(E), \ \r_{\m,\t}^{(i)}(E)$ are the $\n_e$ and
averaged $\n_\m$--$\n_\t$ fluxes evaluated at the point of resonance,
and similarly for the antineutrinos.  We convert the mass-eigenstate
into flavour-eigenstates fluxes using the standard neutrino mixing
matrix~\cite{PDG}.  We adopt the best fit neutrino mixing parameters
reported in~\cite{PDG,Fogli:2008ig,Schwetz:2008er},
\beq
\begin{alignedat}{2}
\D m_{21}^2 &= 7.65 \times 10^{-5} \eV^2, \\
\D m_{32}^2 &= 2.4 \times 10^{-3} \eV^2, \\
\sin^2\th_{12} &= 0.304, \\
\sin^2\th_{23} &= 0.5, \\
\th_{13} &= 0, \\
\d &= 0.
\end{alignedat}
\label{nu mix}
\eeq

\subsubsection{$\n_\m - \n_\t$ oscillations}

Muon and tau neutrinos mix maximally, and oscillate with frequency $f\sub{osc} = \D m_{32}^2/2 E$. The oscillations equidistribute the neutrinos in the two flavour eigenstates, as long as $f\sub{osc}$ is comparable to, or exceeds the interaction rate in the Sun, $\G\sub{int} = \s\sub{tot}(E) N_S (r)$. 
In Fig.~\ref{phase space} we sketch the maximum energy for which 
\beq 
f\sub{osc} \gtrsim \G\sub{int},
\label{osc-int}
\eeq
as a function of the solar radius.  This obtains at optical depth 
\beq
x \gtrsim x\sub{eq}(E) = 1-\pare{\k R_\odot \cal{E} \D m_{32}^2}/2 E^2
\label{xeq}
\eeq
where we used the analytical approximation of \eq{sun}. Although this
condition is not strictly satisfied at all depths for the entire
energy range of interest, it is easy to show that even the
highest-energy neutrinos considered here undergo only a small number
of scatterings $\lesssim 5$, before they lose enough energy and move
out in radius to lower solar density, such that oscillations become
more frequent than their interactions\footnote{On average, every NC
  interaction in the Sun costs neutrinos about 1/2 of their energy (CC interactions of tau neutrinos degrade their energy even more). After $n$
  interactions, the neutrino energy is diminished to $E_n \approx E_0/2^n$, and
  neutrinos have traversed total optical depth $\d x_{\rm{tot},n} =
  \d x_0(2^{n+1}-1)$, where $\d x_0 = \cal{E}/E_0$, according to
  \eq{dx1}. If neutrinos are injected in the centre of the Sun, the
  number of scatterings they undergo till they satisfy \eq{xeq} is
  found by setting $\d x_{\rm{tot},n} = x\sub{eq}(E_n)$. This yields
  $n \approx 4.6 + 3.3\log (E_0/10 \TeV)$. Neutrinos injected away from the
  centre undergo even less scatterings.}.
Since the mediator decays have been chosen to inject equal numbers of
all flavours, we don't rely on the oscillations to establish the
$\nu_\mu$--$\nu_\tau$ average as an initial condition, but only to
maintain it, and thus averaging over the muon and tau neutrino
populations remains an adequate approximation.

\subsection{Propagation \label{sec prop sum}}

We compare the effect of neutrino interactions and oscillations inside
the Sun in Fig.~\ref{phase space}. For all energies in the range $5
\GeV \lesssim E \lesssim 10 \TeV$, resonance between the electron and
muon neutrinos occurs at $r> 0.75 R_\odot$. At this region, less than
$5\%$ of the neutrinos will interact before they exit the Sun,
$P\sub{int} < 0.05$. This allows for separate treatment of the
neutrino interactions in the interior of the Sun and the neutrino
resonant conversions.

We calculate the neutrino signal at the Earth according to the following procedure:
\benu
\renewcommand{\theenumi}{\roman{enumi}}
\item We follow the evolution of the neutrino fluxes from the centre
  of the Sun to $r_c = 0.75 R_\odot$ using \eqs{prop}, taking into
  account the neutrino interactions as described in Sec.~\ref{sec
    inter}.  \eqs{prop} are solved analytically as described below.
  For muon and tau neutrinos we set $\r_{\m,\t}=\r_\m = \r_\t$ (and
  similarly for the antineutrinos), and average over their
  interactions.
We neglect interactions at $r > r_c$, and simply add the contribution
to the neutrino flux from mediator decays between $r_c$ and the
resonance point.

\item We evaluate the level crossings at resonance, using
  \eqs{flux1,2,3}. We convert the mass-eigenstate fluxes back into
  flavour-eigenstate ones using the standard mixing parameters given
  in \eq{nu mix}.

\item We add the neutrino flux generated by decays between the point
  of resonance and the Earth.

\eenu
We shall now describe these steps in more detail.

\bigskip

\begin{figure}[t]
\centering
\includegraphics[width=0.6\linewidth]{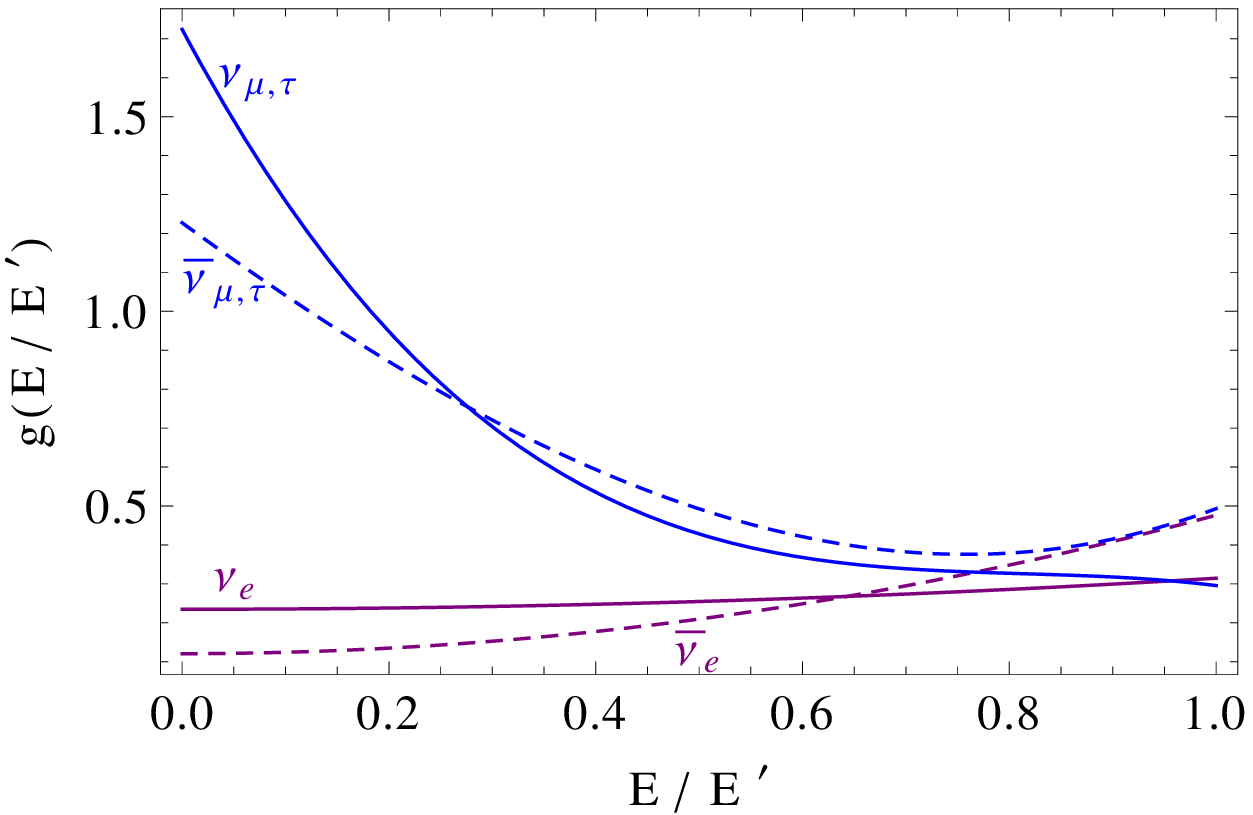}
\caption{The regeneration functions of \eqs{regen f}, for $\n_e$ (purple, solid), $\bar{\n}_e$ (purple, dashed), $\n_{\m,\t}$ (blue, solid), $\bar{\n}_{\m,\t}$ (blue, dashed).}
\label{regen fun}
\end{figure}

The evolution equations \eqref{prop} take the form
\beq
\frac{\pd \r_j}{\pd x} = h(x,E) + \frac{1}{\cal{E}} \left[-E \r_j(x,E) + \int_E^\infty  dE' g_j\pare{E/E'} \r_j(x,E')\right],
\label{rho-diff}
\eeq
where $\cal{E}$ is given by \eqs{cal E} and \eqref{cal E num}.
For $r<r_c$, the exponential approximation of \eq{sun} for the solar profile is valid and the neutrino injection term $h(x,E)$ is expressed analytically by
\beq
h_a(x,E) =   \frac{C_\odot \: \rm{BR}_\n}{\b m_\x} \: \frac{\k R_\odot}{\b\g\t} (1-x)^{\frac{\k R_\odot}{\b\g\t}-1},
\label{ha}
\eeq
for the energy interval of \eq{E nu}.
The regeneration functions appearing in \eqs{rho-diff} are
\beq
\begin{alignedat}{2}
g_e (u) &= \frac{a_\n + b_\n u^2}{a_\n + b_\n/3 + c_\n},   \\
g_{\m,\t} (u) &= \frac{a_\n + b_\n u^2 + c_\n/2 \: f_{\t \rightarrow \t}(u)}{a_\n + b_\n/3 + c_\n}, \\
g_{\bar{e}} (u) &= \frac{a_{\bar{\n}} + b_{\bar{\n}} u^2}{a_{\bar{\n}} + b_{\bar{\n}}/3 + c_{\bar{\n}}},  \\
g_{\bar{\m},\bar{\t}} (u) &= \frac{a_{\bar{\n}} + b_{\bar{\n}} u^2 + c_{\bar{\n}}/2 \: f_{\bar{\t} \rightarrow \bar{\t}}(u)}{a_{\bar{\n}} + b_{\bar{\n}}/3 + c_{\bar{\n}}},
\end{alignedat}
\label{regen f}
\eeq
and are sketched in Fig.~\ref{regen fun}.

Equation \eqref{rho-diff} can be solved analytically for $g=0$ and $g=1$. For a general function $g$, it is possible to solve \eq{rho-diff} perturbatively around these two cases. The range of values of the regeneration functions, presented in Fig.~\ref{regen fun}, renders this a suitable treatment for the neutrino propagation in the Sun.
For the $\n_e$ and $\bar{\n}_e$ fluxes, $g_e, \ g_{\bar{e}} < 1$, and the perturbative solution around $g=0$ is warranted.
For the $\n_{\m,\t}$ and $\bar{\n}_{\m,\t}$ fluxes, the perturbative solution around $g=1$ appears formally more appropriate. We have implemented both solutions and checked that, for the DM masses and mediator lifetimes considered here, the differences in the final spectra are small and do not qualitatively change our conclusions. For simplicity, we will thus adopt the perturbative solution around $g=0$ for all fluxes.

The exact solution of \eq{rho-diff} for $g=0$ is
\beq
\r_0(x,E) = \int_0^x dx' \: e^{-\frac{E (x-x')}{\cal{E}}} h(x',E).
\label{rho0}
\eeq
Let
\beq
\r(x,E) = \r_0(x,E) + \r_1(x,E) \label{pert},
\eeq
where $\r_1(x,E)$ obeys, to 1st order in $g$, the differential equation
\beq
\frac{\pd \r_1}{\pd x} = h_1(x,E) - \frac{E}{\cal{E}} \r_1(x,E),
\label{rho1-diff}
\eeq
with
\beq
h_1(x,E) \equiv \frac{1}{\cal{E}} \int_E^\infty dE' \: g(E/E') \: \r_0(x,E').
\label{h1}
\eeq
\eq{rho1-diff} has the same form as \eq{rho-diff} for $g=0$, and its solution is
\beq
\r_1(x,E) = \int_0^x dx' \: e^{-\frac{E (x-x')}{\cal{E}}} h_1(x',E)  \label{rho1}
\eeq
\eqs{rho0}, \eqref{pert}, \eqref{h1} and \eqref{rho1} yield the perturbative solution to the evolution equation, to first order in $g$.

%\medskip

The contribution to the neutrino flux from decays of the mediators between $r_c$ and $r\sub{res}(E)$ is
\beq
\d\r\ssub{I}(E) = \frac{C_\odot \: \rm{BR}_\n}{\b m_\x} \sqpare{ \exp\pare{-\frac{r_c}{\b\g\t}} - \exp\pare{-\frac{r\sub{res}(E)}{\b\g\t}}}.
\label{Drho I}
\eeq
For $\g \t \gtrsim 2 \snd$, the neutrino production beyond the point of resonance is significant and is given by 
\beq
\d\r\ssub{II}(E) = \frac{C_\odot \: \rm{BR}_\n}{\b m_\x} \sqpare{\exp\pare{-\frac{r\sub{res}(E)}{\b\g\t}} -  \exp\pare{-\frac{R\sub{SE}}{\b\g\t}}},
\label{Drho II}
\eeq
with
$R\sub{SE} = 1 \AU$

\bigskip

\begin{figure}[t]
\centering
\includegraphics[width=\linewidth]{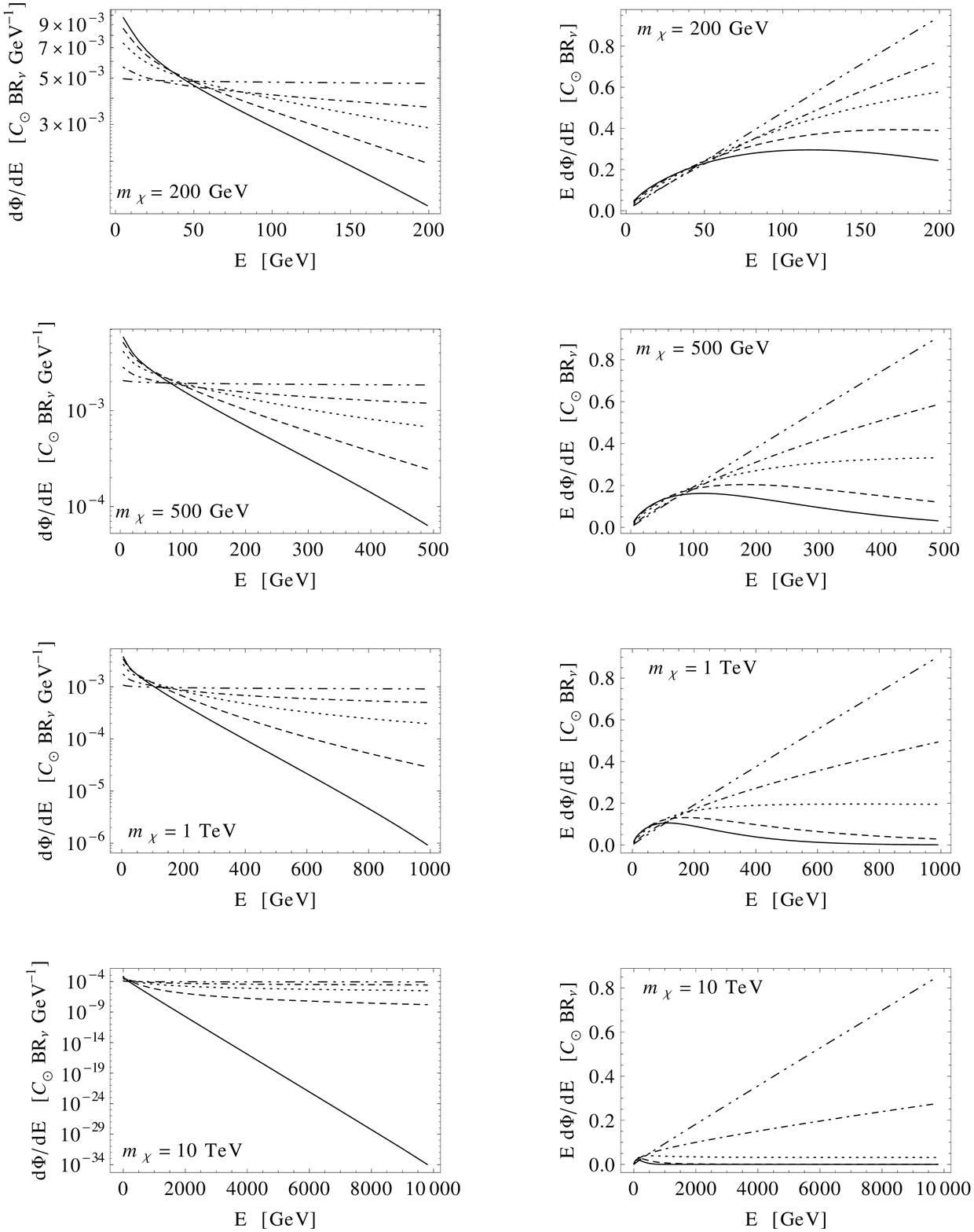}
(a) \hspace{0.5\linewidth} (b)
\caption{$\n_e$ fluxes for $\g\t =$ 0.001~s (solid), 0.1~s (dashed) 0.3~s (dotted) 1~s (dot-dashed), 10~s (dot-dot-dashed), and DM masses $m_\x =$ 200~GeV (1st row), 500~GeV (2nd row), 1~TeV (3rd row), 10~TeV (4th row).}  
\label{All tau e}
\end{figure}

\begin{figure}[t]
\includegraphics[width=0.95\linewidth]{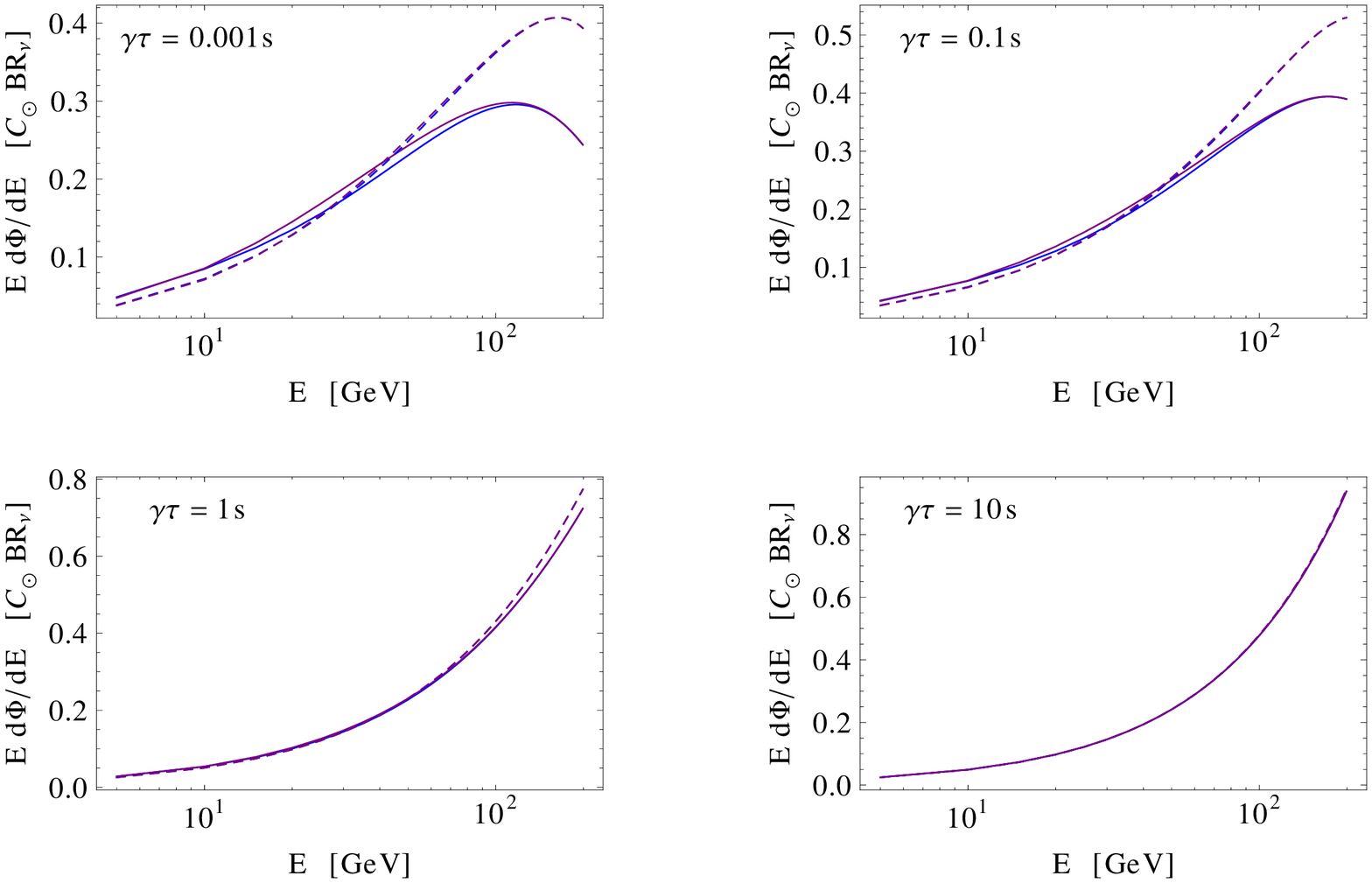}
\caption{Neutrino fluxes at the Earth for DM mass $m_\x = 200 \GeV$, and mediator lifetimes $\g\t =$ 0.001~s, 0.1~s, 1~s, 10~s.  In each graph the lines correspond to:
$\n_e$ flux (blue, solid), $\n_{\m,\t}$ flux (purple, solid), $\bar{\n}_e$ flux (blue, dashed), and $\bar{\n}_{\m,\t}$ flux (purple, dashed).}
\label{200GeV}
\bigskip
\bigskip
\includegraphics[width=0.95\linewidth]{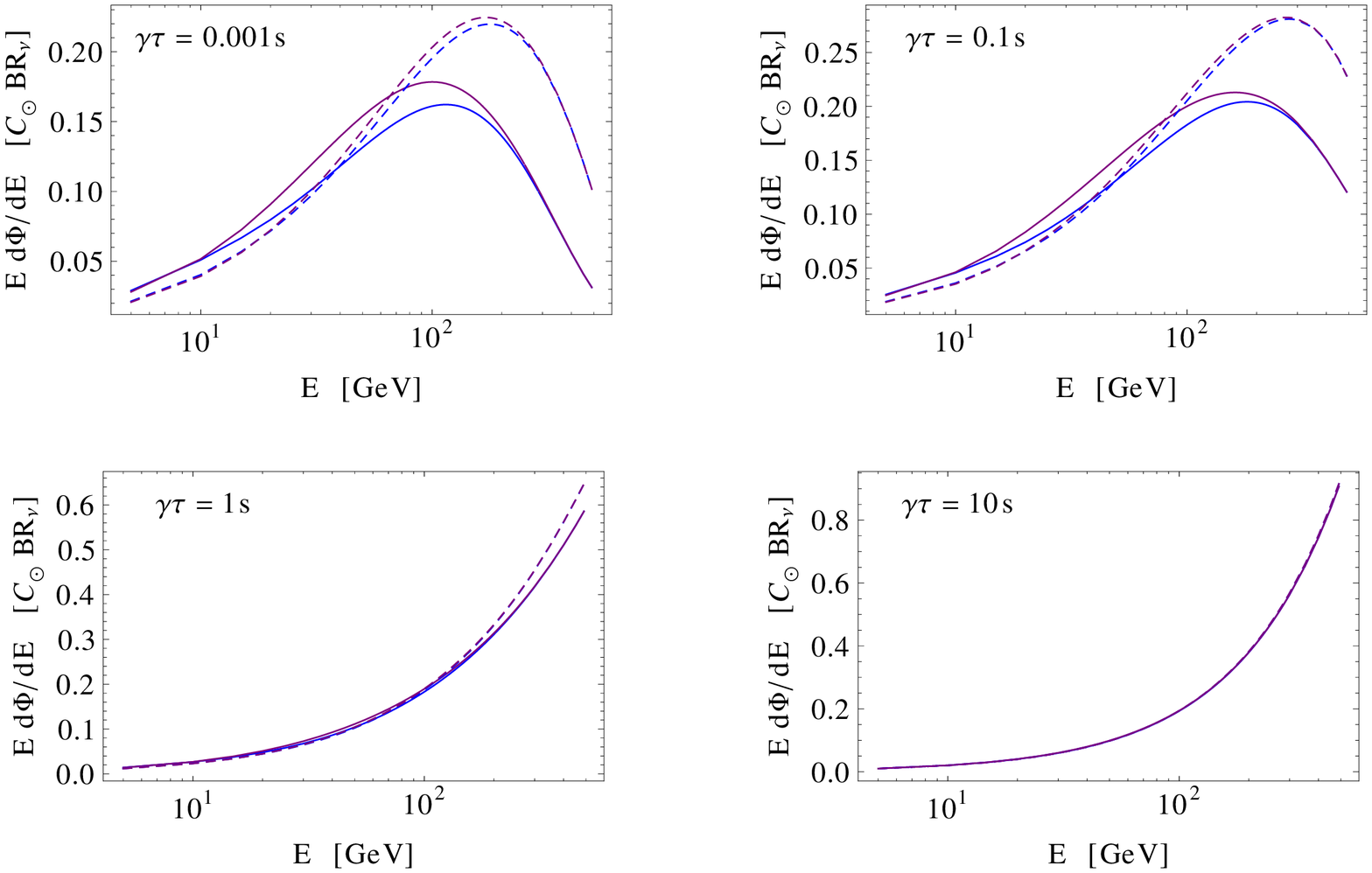}
\caption{Same as in Fig.~\ref{200GeV}, for DM mass $m_\x = 500 \GeV$.}
\label{500GeV}
\end{figure}

\begin{figure}[t]
\includegraphics[width=0.95\linewidth]{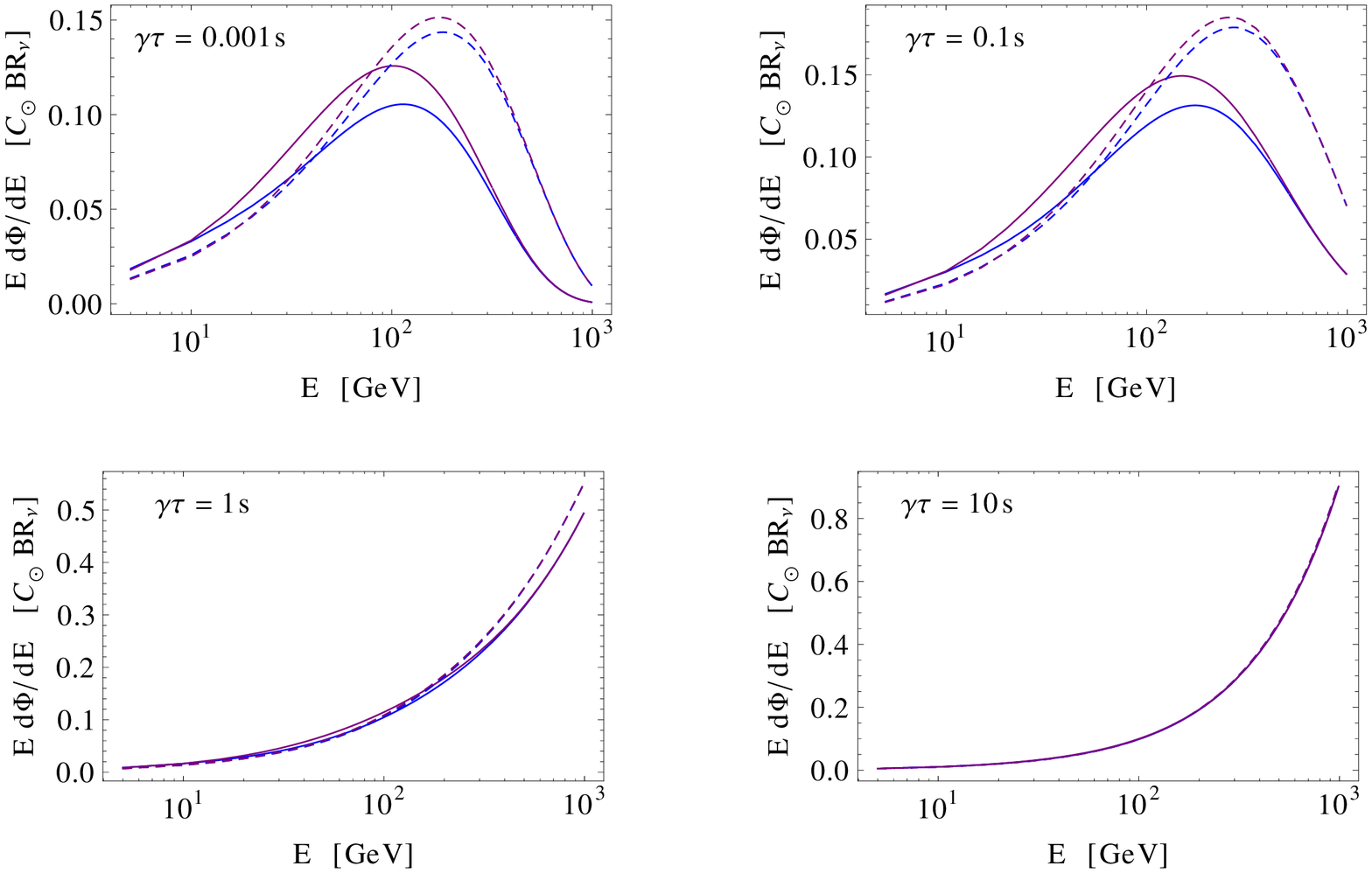}
\caption{Same as in Fig.~\ref{200GeV}, for DM mass $m_\x = 1 \TeV$.}
\label{1TeV}
\bigskip
\bigskip
\bigskip
\includegraphics[width=0.95\linewidth]{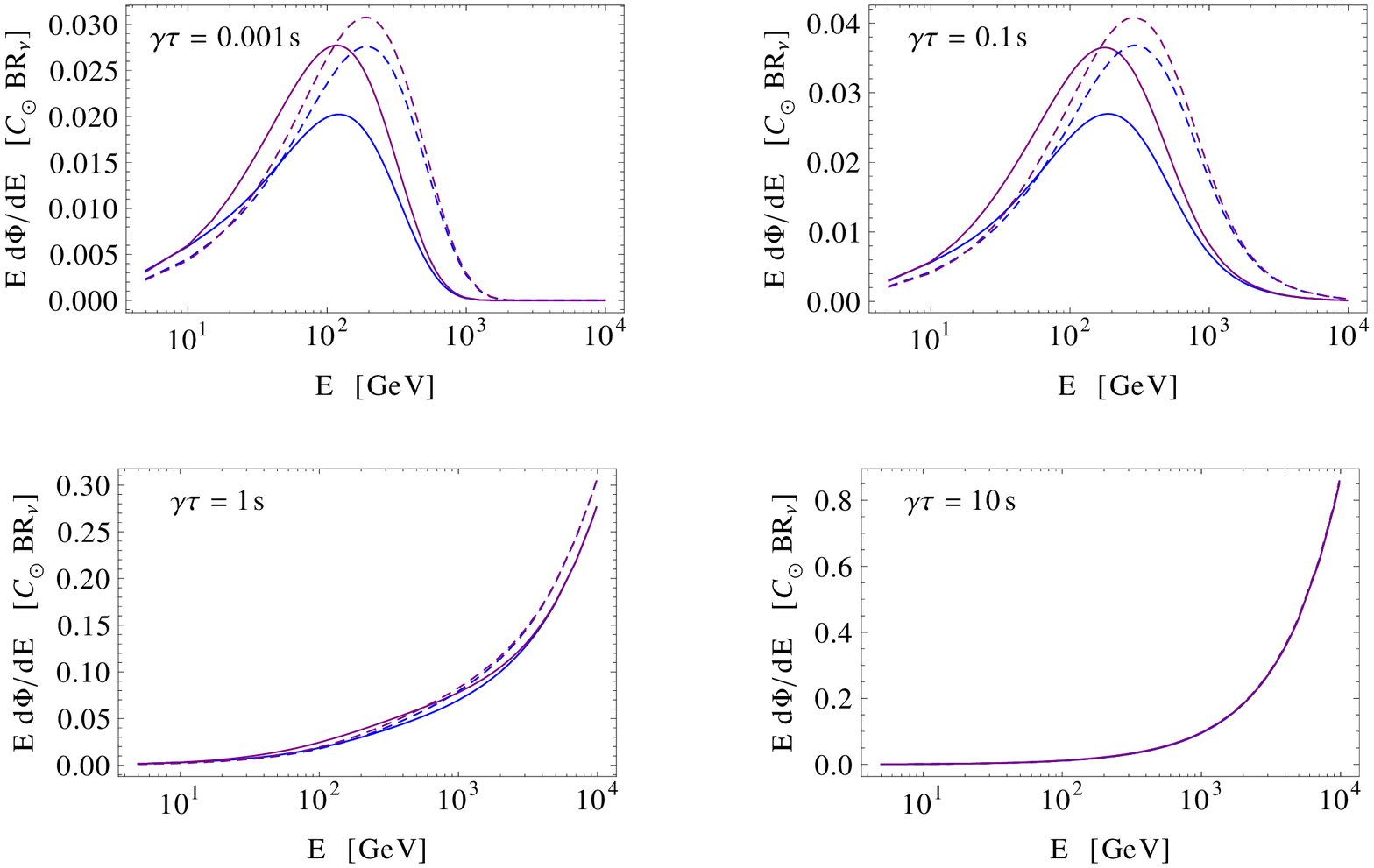}
\caption{Same as in Fig.~\ref{200GeV}, for DM mass $m_\x = 10 \TeV$.}
\label{10TeV}
\end{figure}

The final neutrino fluxes at Earth, calculated as outlined above, are
presented in Figs.~\ref{All tau e} -- \ref{10TeV} for various choices
of the DM mass and mediator lifetime, and discussed in the following section.

%%%%%%%%%%%%%%%%%%%%%%%%%%%%%%%%%%%%%
%%%%%%%%%%%%%%%%%%%%%%%%%%%%%%%%%%%%%

%%%%%%%%%%%%%%%%%%%%%%%%%%%%%%%%%%%%%
%%%%%%%%%%%%%%%%%%%%%%%%%%%%%%%%%%%%%

%%%%%%%%%%%%%%%%%%%%%%%%%%%%%%%%%%%%%

%%%%%%%%%%%%%%%%%%%%%%%%%%%%%%%%%%%%%%

%%%%%%%%%%%%%%%%%%%%%%%%%%%%%%%%%%%%%%%%%%%%%%%%%%%%%%%%%%%%%%%%%%%%
%%%%%%%%%%%%%%%%%%%%%%%%%%%%%%%%%%%%%%%%%%%%%%%%%%%%%%%%%%%%%%%%%%%%

\section{Neutrino signals \label{sec signal}}

\subsection{Neutrino flux results}

In Figs.~\ref{All tau e} -- \ref{10TeV} we present the final differential neutrino
fluxes at Earth, $d\F/dE = \r(1 \rm{A.U.}, E)$, calculated as outlined in previous section, for
various choices of the DM mass and mediator lifetime.  All fluxes are
normalised to the DM capture rate in the Sun, $C_\odot$, and the
branching ratio $\rm{BR}_\n$ for decay to neutrinos.

In Fig.~\ref{All tau e}, we compare the $\n_e$ fluxes obtained for
different mediator lifetimes, $\g\t = 0.001 \snd, \ 0.1 \snd, \ 0.3 \snd, \ 1
\snd, \ 10 \snd$ (always assuming $\g \gg 1$).  
For small mediator lifetime, absorption effects result in an exponential
suppression of the flux at high energy, while down scattering or
regeneration effects cause a pile up of neutrinos at low energy.
These effects can both clearly be seen in the left-hand panels of 
Fig.~\ref{All tau e}.
Relativistic particles transverse the entire radius of the Sun in
$\sim 2.2 \snd$, while the core of the Sun corresponds to $\sim 1/10$
of this distance (cf. \eq{sun}). Thus, the smallest mediator lifetime
we consider, $\g\t = 0.001 \snd$, effectively corresponds to the
standard scenario of DM annihilation directly into SM particles in the
centre of the Sun\footnote{We note that the neutrino spectrum
  resulting from the prompt decay ($\g\t \lesssim 0.2 \snd$, with $\g
  \gg 1$) of each metastable mediator to a neutrino pair resembles the
  neutrino spectrum obtained from DM annihilation to relativistic SM
  particles such as $\t^+\t^-$, rather than that for DM annihilation
  directly to a neutrino pair. In the latter case, neutrinos are
  emitted monoenergetically, which results in a residual peak in the
  final spectrum, at the maximum energy. In the case of decays
  of boosted particles, neutrinos are injected with an extended
  spectrum, as described in \eq{E nu}.}.
For the largest lifetime we consider, $\g\t = 10 \snd$, almost all the
mediators decay outside the Sun, so absorption is negligible and the
characteristic flat injection spectrum of boosted 2-body decays
(discussed in Sec.~\ref{sec inj}) is reproduced.  Between these two
limiting cases, the fluxes depend sensitively on the mediator
lifetime.

We can quantify the degree of attenuation in terms of the neutrino
injection radius as follows: if we inject all neutrinos at radius
$r=r\sub{inj}$ and consider only absorption effects, the solution of the 
flux evolution equation (\eq{rho-diff}) is simply
\beq
\r = \r\sub{inj} e^{-\frac{E}{\cal{E}} \D x},
\label{rho-simple}
\eeq 
where $\r\sub{inj}$ is the flux at the injection radius, and $\D x$ is the 
optical depth traversed. 
Neutrinos injected at $r=r\sub{inj}$ traverse an optical depth of 
$\D x = \exp\pare{-\frac{r\sub{inj}}{\k R_\odot}}$ as they propagate to
$r=R_\odot$ (see \eq{x,r anal}).
For $r\sub{inj}=0$, $\D x \simeq 1$, and thus $\r=\r\sub{inj}
e^{-\frac{E}{\cal{E}}}$.  More generally, if we define a critical
energy ${\cal{E}'}$ as
\beq
{\cal{E'}}(r\sub{inj})= {\cal{E}}\exp\left(\frac{r\sub{inj}}{\k R_\odot}\right),
\label{E'}
\eeq
the solution for the attenuated flux at the solar surface can be
expressed as
\beq
\r  = \r\sub{inj} e^{-\frac{E}{\cal{E'}}}.
\label{exp}
\eeq
The critical energy ${\cal{E'}}$ controls the exponential suppression
of the flux at high energies.  Since ${\cal{E'}}$ grows exponentially
with injection radius, we see that this critical energy for absorption
grows rapidly as soon as the injection radius is outside the core,
$r\sub{inj} \gtrsim \k R_\odot$.

In the scenario considered here, mediator decays inject neutrinos not at
a single point but with a distribution of radii. To account for the
absorption we must employ the injection function of \eq{inj}, or equivalently
\beq
%\left.\frac{d\F}{dE}\right|\sub{E \gtrsim \cal{E}}  \propto    
%\frac{1}{\b\g\t} \int_0^\infty  \exp \pare{-\frac{r}{\b\g\t}} \exp\pare{-\frac{E}{\cal{E'}(r)}}  dr.
% OR
\left.\frac{d\F}{dE}\right|\sub{E \gtrsim \cal{E}} =  \frac{C_\odot \: \rm{BR_\n}}{\b m_\x} 
\frac{1}{\b\g\t} \int_0^\infty  \exp \pare{-\frac{r}{\b\g\t}} \exp\pare{-\frac{E}{\cal{E'}(r)}}  dr.
\label{approx}
\eeq 
This provides a good approximation for the flux for high energies $E \gtrsim \cal{E}$,
where absorption is the leading effect.  Regeneration effects become
relevant at lower energies, where a pile up of low energy neutrinos
enhances the fluxes above the levels indicated by \eq{approx}.

Annihilation via the mediators enhances the prospects for detection of
a DM annihilation signal from the Sun.  In the right-hand panels of
Fig.~\ref{All tau e} (and in Figs.~\ref{200GeV} -- \ref{10TeV}) we
plot the quantity $Ed\F/dE$.  As the neutrino scattering cross-section
on nucleons is proportional to the energy (see \eq{sigma tot}), the
area under the $Ed\F/dE \ vs. \ E$ graphs is a guide to the relative
event rates in neutrino detector.
For all DM masses and mediator lifetimes, the prospects for detection
are better at neutrino energies $E_\n \gtrsim 100 \GeV$, and
dramatically so as the DM mass is increased.
Since the high energy neutrinos are the most enhanced and also have
the largest detection cross sections, the mediator scenario is
particularly effective in enhancing event rates.

The neutrino spectral shape is a striking signature of DM annihilation
via metastable states.  In the standard scenario, there is never
significant neutrino flux at energies $E_\n \gtrsim 1 \TeV$.  If DM
annihilates directly into SM particles, $E d\F/dE$ exhibits a peak
around $\sim 100 \GeV$ (see Figs.~\ref{200GeV} -- \ref{10TeV}).  The
peak occurs because the absorption increases strongly with energy and
always quells the spectrum at higher energies.  In fact, for $m_\x >
1\TeV$, neutrino interactions in the Sun drive the spectrum to a fixed
shape independent of the DM mass or annihilation channel (dubbed the
``limit spectrum'' in Ref.~\cite{Cirelli:2005gh}) with a sharp cutoff
at $E_\n \approx 1 \TeV$.  These features are reproduced by our
calculations for $\g\t = 0.001 \snd$.
In the presence of mediators, however, the peak of the spectrum moves
to higher energy if $\g\t$ is small, and completely disappears for
$\g\t \gtrsim 0.3 \snd$ (which corresponds to mediator decay outside
the core of the Sun).  Although the small shift in the spectral shape for small $\g\t$
may not be readily discernible, the rising trend of the spectrum
($E d\F/dE$) at high energies that occurs in the case of sufficiently
long-lived mediators is a firm signature of this scenario. It cannot
arise in the standard scenario, even for very large DM mass.  A
potential observation of a neutrino signal from the Sun at energies
$E_\n \gtrsim 1 \TeV$ can thus be justified only in the context of DM
annihilation into metastable states with $\g\t > 0.1 \snd$.
%, as exhibited in Fig.~\ref{All tau e}(b).

We compare the fluxes of all neutrinos flavours, for both neutrinos
and antineutrinos, in Figs.~\ref{200GeV} -- \ref{10TeV}.  We have
assumed that the mediators decay with equal branching ratio to all
flavours.  However, interactions induce flux differences between
neutrinos and antineutrinos, and between the different flavours.
Neutrinos interact more strongly in the Sun than antineutrinos, thus
are absorbed more, resulting in smaller fluxes.  Due to the $\n_\t$
and $\bar{\n}_\t$ regeneration, the muon/tau neutrino and antineutrino
fluxes are higher than the electron-type ones. These features are more
apparent for larger DM masses and smaller mediator lifetimes, when
interactions are stronger.  As the mediator lifetime is increased,
such that interactions become less important, the spectra for all
flavours of $\nu$ and $\bar\nu$ merge.

\subsection{Muon event rates}

We shall now estimate the number of muon track events in the IceCube
detector for a range of mediator lifetimes, relative to the scenario
in which the neutrinos are produced in the solar core.  We estimate
event rates following the procedure in
Refs.~\cite{Gandhi:1995tf,Dutta:2000hh,Beacom:2003nh}, which is 
outlined below.

A muon neutrino of energy $E$ produces a muon with energy $E_\mu = E
(1-y)$ when it undergoes a charged current interaction, where $y$ is
the charged current inelasticity parameter.  For simplicity we will
adopt the mean values of $y$ tabulated in Ref.~\cite{Gandhi:1995tf},
which are approximately $\sim 0.45$ for neutrinos and $\sim 0.35$ for
antineutrinos.
Muons propagate over a long distance before they decay, such that
IceCube can detect muons created from $\nu_\mu$ interactions well
outside the detector.  The muons lose energy as they propagate, and
have a range given by
\begin{equation}
R_\mu(E_\mu, E_\mu^\textrm{thr}) 
= \frac{1}{\beta} 
\ln \left[ \frac{\alpha+\beta E_\mu}{\alpha + \beta E_\mu^\textrm{thr}} \right],
\end{equation}
where $\alpha = 2.0 \MeV \cm^2/\rm{g}$, $\beta = 4.2 \times
10^{-6}\cm^2 / \rm{g}$, and $E_\mu^\textrm{thr}$ is the muon energy
detection threshold for the detector (see, e.g.,
Refs.~\cite{Gandhi:1995tf,Dutta:2000hh,Groom:2001kq,Beacom:2003nh}).
We will take $E_\mu^\textrm{thr}$ to be 100 GeV.
The probability that a neutrino of energy $E$ makes a muon track that
is detected with energy above $E_\mu^\textrm{thr}$ is then
\begin{equation}
P(E,E_\mu^\textrm{thr}) = \rho_N  N_A  \sigma\sub{nucleon}  R_\mu(E(1-y), E_\mu^\textrm{thr}),
\label{eq:P}
\end{equation}
where $\rho_N$ is the target nucleon density and $N_A$ is Avogadro's number. 
The muon track event rate is thus proportional to
\begin{equation}
\int \frac{ d\F(E)}{dE} P(E,E_\mu^\textrm{thr}) A^\textrm{eff} dE
\end{equation}
where $A^\textrm{eff}$ is the effective area of the detector, which
is $\sim$1 km$^2$ for IceCube~\cite{Ahrens:2003ix}.

In table~\ref{fluxes}, we show the relative number of muon track
events for different choices of the mediator lifetime, normalised to
the case where $\tau\simeq 0$ (i.e., production in the centre of the
Sun). For the latter, we adopt the fluxes obtained here for $\g\t = 0.001 \snd$, 
in order to retain consistency with our calculations.  
Both neutrino and antineutrino events are included, weighted
according to their respective CC interaction probabilities (which have
a ratio of approximately 2:1 in ice).  For dark matter masses close
the transparency energy of the Sun $\sim$ 200 GeV, the effect of the
metastable mediator is to enhance the flux by about a factor 2.  For
larger dark matter masses, well above the energy at which the Sun
becomes opaque neutrinos, annihilation via metastable mediators has a
huge effect on the flux, with an enhancement of an order of magnitude
for lifetime as short as 0.1s, and an enhancement of about 1000 for
lifetime comparable to the solar radius.  Particularly for these heavy
dark matter masses, the presence of the metastable mediator greatly
enhances the flux which emerges from the Sun, and therefore the
detection prospects.

\begin{table}[t]
\begin{center}
\begin{tabular}{| c || c | c | c | c | }
\hline
%$m_\x$ & 200 GeV & 500 GeV & 1 TeV& 10 TeV \\ \hline
%$\g \t$ & \multicolumn{4}{|c|}{ $\F(\g\t) / \F(\g\t \rightarrow 0)$ for $E_\m > 100 \GeV$ }   \\ \hline
        & 200 GeV & 500 GeV & 1 TeV& 10 TeV \\ \hline \hline
0.1 s   & 1.4 & 1.9 & 2.9   & $1.2 \times 10$ \\ \hline
0.3 s   & 1.9 & 3.4 & 7.6   & $2.7 \times 10^2$  \\ \hline
1s      & 2.3 & 4.8 & $1.4 \times 10$  & $1.4 \times 10^3$ \\ \hline
10 s    & 2.9 & 6.7 & $2.2 \times 10$  & $3.6 \times 10^3$ \\ \hline
\end{tabular}
\end{center}
\caption{The ratio of $\m^\pm$ events observable in the presence of
  mediators of lifetime $\g\t$ over those in the absence of metastable
  mediators, $\F(\g\t) / \F (\g\t \rightarrow 0)$, integrated over
  muon energies $E_\m > 100 \GeV$, for various dark-matter masses and
  mediator lifetimes. We have used the calculated fluxes from our
  analysis for $\g\t = 0.001 \snd$ as the benchmark fluxes for DM
  annihilation directly into SM particles.}
\label{fluxes}
\end{table}

%%%%%%%%%%%%%%%%%%%%%%%%%%%%%%%%%%%%%%%%%%%%%%%%%%%%%%%%%%%%%%%%%%%%
%%%%%%%%%%%%%%%%%%%%%%%%%%%%%%%%%%%%%%%%%%%%%%%%%%%%%%%%%%%%%%%%%%%%

\section{Conclusions \label{concl}}

The search for high energy neutrinos produced by the annihilation of DM
captured in the Sun is a promising means of indirect DM detection.
A future detection can provide information about the DM annihilation
channels, which is critical for the identification of the particle
nature of DM.
It is possible, however, that DM does not annihilate directly into SM
particles. We have examined a scenario in which DM annihilates to
metastable mediators, which subsequently decay to SM states.  In these
models, though the DM and SM sectors are secluded from each other and
communicate only via the mediators, the standard thermal-WIMP scenario
can be realised as usual.  Existing DM indirect detection bounds are
modified in these models.  Particularly interesting is the signal from
DM capture in the Sun, which is enhanced at high energy and provides
distinct features which can unmask the secluded nature of DM.

In the case where the mediators propagate out from the dense solar
core before decaying, the neutrinos produced in the decays experience
a much smaller optical depth and thus much less absorption, compared
to neutrino production at the centre of the Sun.  In the standard
scenario, the absorption is so severe that the flux is almost
completely suppressed beyond about 1 TeV.  The observation of a solar
neutrino flux beyond 1 TeV would thus be striking indication of the
presence of metastable mediators.
These features would be discernible even with small number of events,
or poor detector energy resolution, and remain valid independent of
the specific decay channel of the mediators into SM states.
Finally, this enhancement of the neutrino flux at high energy is
particularly effective in improving detection prospects, given that
the detection cross sections grow with energy.

%%%%%%%%%%%%%%%%%%%%%%%%%%%%%%%%%%%%%%%%%%%%%%%%%%%%%%%%%%%%%%%%%%%%
%%%%%%%%%%%%%%%%%%%%%%%%%%%%%%%%%%%%%%%%%%%%%%%%%%%%%%%%%%%%%%%%%%%%

%\section*{Acknowledgements} 
\acknowledgments 
We thank John Beacom for helpful discussions.  
This work was supported, in part, by the Australian Research Council.

\bibliographystyle{JHEP}
\providecommand{\href}[2]{#2}\begingroup\raggedright\endgroup

\end{document}